\documentclass[10pt]{article}
\usepackage{epsfig}
\usepackage{graphics}
\usepackage[english]{babel}      

\newcommand{\be}{\begin{equation}}
\newcommand{\ee}{\end{equation}}
\newcommand{\ba}{\begin{eqnarray}}
\newcommand{\ea}{\end{eqnarray}}

\begin{document}
\title{On Contributions of Two-Particle Correlation Functions 
to the Intermittent Behavior in Heavy-Ion Collisions}
\author{Abdelnasser M. Tawfik\\ \\
{\normalsize\sf FB Physik, Marburg University, D-35032 Marburg, Germany} 
}

\date{}
\maketitle
\parindent5mm

\renewcommand{\thefootnote}{\arabic{footnote}}

\begin{abstract}
  The factorial moments (FM) of the multiplicity distributions in
  one- and two-dimensions are studied for Pb+Pb collisions at
  $158$~AGeV. Obvious intermittence slopes are observed in the
  relation between FM, and the partition number $M$. The
  Bose--Einstein correlations (BEC) are investigated in the emulsion
  data sample. The influences of the two-particle correlation functions
  (TPCF) on the scaling behavior of FM are also investigated. The
  dependence of TPCF on the invariant relative momentum $Q$, as well
  as on the pseudo-rapidity, $\delta\eta$, are calculated. It is found
  that TPCF fulfill the requirements of the power-law scaling comparable
  with the scaling rule of FM. The expected TPCF contributions to FM
  are quantitatively estimated.\\ \\

{\bf Keywords:} fluctuations, intermittence, Bose--Einstein correlations,
many--particle correlation functions, strip integral correlations,
heavy--ion collisions, quark--gluon plasma\\ \\

{\bf Pacs:} 25.75.G, 61.43, 12.38.M, 64.60.Q
\end{abstract}

\section{Introduction} \label{bec_sec1}

The intensity interference between the produced particles is the
unique experimental tool to directly probe the space-time evolution of
the interacting system. First, it was used in the radio-astronomy (HBT)
\cite{ba2} to estimate the astronomical distances. After that, the
principles of HBT are used by {G. Goldhaber, S. Goldhaber, W.~Lee
  and A. Pais} (GGLP), in order to study the angular distribution of
produced particles in $p-\bar{p}$ annihilation \cite{ba3}. GGLP
describes the phenomenon, that the like-sign charged (identical)
particles are emitted within space angles smaller than that of
unlike-sign charged (non-identical) ones. Evidently, this appears as a
reason of the symmetrical properties of boson wave functions and
therefore, the emission probability is very sensitive to the {\it
Bose--Einstein statistics}. GGLP is therefore known as Bose--Einstein
correlations (BEC). In this article, we consider the simplest case of
two-particle correlation functions (TPCF), namely BEC. It is known
that the shape of TPCF enhancement depending on the relative momentum
$Q$ can be related to the emission source dimension $R$ \cite{ba3}
and/or to it's life time $\tau$ \cite{ba4}. In principle, the
interference between identical produced particles occurs and
consequently their TPCF get powerful, when certain conditions are
satisfactorily fulfilled. The produced particles should be
non-coherent and simultaneously emitted, i.e.\ the following similar to
the Heisenberg's uncertainty relation has to be satisfied. $\Delta p
\cdot \Delta r \le 2 \pi \hbar$ \cite{ba2,ba3,ba5,ba6}. $\Delta
p=|p_1-p_2|$, and $\Delta r=|r_1-r_2|$ are the relative momentum and
distance of the considered particle pair, respectively. In heavy-ion
collisions the typical dimension of the emission source is $\Delta
r\approx 6$~fm \cite{Alt1}. Then the interference takes place for
$\Delta p$ up to $50$~MeV/c \cite{ba7}. As mentioned above, TPCF can
be used to directly measure many important parameters (e.g.\ dimension,
lifetime, expansion velocity of the emission source). These
measurements are highly necessary to study the dynamics of strong
interactions and also to probe whether the energy density exceeds the
critical value to produce the quark-gluon plasma (QGP) \cite{ba8,ba9}.

The two-particle interferometry is recently used to study the
dynamics of emission source until the final state of particle
production \cite{ba11}. The Fourier transformation of the
space-time density distribution of emission source can
experimentally be estimated by TPCF. Generally, the intensity
interference is not a characteristic phenomenon of bosons alone.
It can also be applied for the fermions. Of course, for fermions, the
intensity correlations appear due to the decreased likelihood.
Furthermore, the two-proton interferometry introduced in
\cite{ba15} is an effective applicable tool at low energy. Since
the protons, in contrast to the pions, were present and
effectively participating in all phases of the colliding system
from the beginning until the {\it freeze-out}. The suggestion of
the particle--antiparticle interferometry \cite{ba16} is generally
based on the discrepancies between the s-channels of
the particle--antiparticle, and that of the particle--particle interactions.

Coming to the second topic of this article; the factorial
moments (FM) which have found increasing interest among the high
energy physicists, since the pioneer works of Bai{\l}as and
Peschanski during the second half of 1980's \cite{ba30}. The
concept {``intermittence"} is evidently used to describe the
fluctuations in the density distributions of particle
production, i.e.\ the power law of FM as functions of the successive
phase-space partitions $\delta\eta\rightarrow 0$. To get a
satisfactory explanation for such power-scaling behavior, many
models are proposed. All these theoretical approaches are trying
to give an answer to the question: {``What are the possible
sources of the power-law behavior in the multi-particle
production?''} \cite{TawPress1}. Following is a short list of
some of these approaches.
\begin{itemize}
\item {Self-similar branching processes,}\vspace*{-4pt}
\item {Conventional short-range correlations,}\vspace*{-4pt}
\item {Multi-particle cascade at the final state,}\vspace*{-4pt}
\item {Critical second-order phase transition to QGP.}
\end{itemize}
Besides these dynamical sources, there are some features
\cite{ba18,ba20,ba21} for the responsibility of TPCF for the
enhanced particle fluctuations, especially in small phase-space
intervals. Through successive decreasing the relative momentum
of two identical particles, an exponential increasing of their TPCF is high
probably expected. Here we are trying to ascertain whether the
reduction of the phase-space of pseudo-rapidity $\Delta\eta$,
which basically are performed to investigate FM,
conducts to auxiliary enhancement of TPCF or {vice
versa}. Experimentally, quite strong evidences are registered
\cite{ba21} for the increment of TPCF with successive reduction
of relative momentum $Q=|p_1-p_2|$, as well as with the
decreasing of
the pseudo-rapidity phase-space $\delta\eta=|\eta_1-\eta_2|$.

In this paper both subjects and their traces in  Pb+Pb
collisions at $158$~AGeV are researched. For the data
acquisition and the experimental setup we refer to \cite{ba241,ba242} and
also to the next section. Here it is important to emphasize
that by using the nuclear emulsion one has on the one hand angular
resolution and local accuracy as good as $\sim$$1$~mrad and
$1\; \mu$m, respectively. But on the other hand the identification of
produced particles and the estimation of their charges and
impulses can not be achieved. This article is organized as follows:
Section \ref{bec_sec2} introduces the experimental setup. 
Section \ref{bec_sec3} deals with the basic definitions and
formalism of TPCF, FM, and the strip integral correlations.
Section \ref{bec_sec4} is devoted to the experimental results. Finally, the
summary and final conclusions are given in Section \ref{bec_sec5}.

\section{Experimental Setup}
\label{bec_sec2}

The data used in this paper are retrieved from some of Pb chambers
irradiated at CERN-SPS during 1996 for the EMU01 Collaboration.
This material is consummately measured and analyzed at Marburg
university by using our measuring system MIRACLE Lab
\cite{ba241,ba242,TawDs}. The collisions are recorded with
$^{82}$Pb beam accelerated to an incident momentum of 158
AGeV/c and directed towards stationary lead foil with thickness $\sim$
250 $\mu$m. The lead foil was positioned in the front of seven
plastic sheets coated on both sides by the nuclear emulsion
FUJIET-7B. The emulsion is acting in these investigations as
detector. The emulsion has a constant area ($10\times
10$~cm$^2$) but different depths. The first two sheets have emulsions with
180 $\mu$m in depth. The rest are coated by emulsions with thickness
$90 \;\mu$m.

The exposure process is mainly controlled by counting the heavily
ionizing particles by scintillator and discriminator with relatively
high threshold settings. The number of beam particles is counted by
using an additional counter installed behind the chamber.  Scaler and
driver electronics (CAEN N145) are also used to build up and to transfer
pulses to the SPS {``kicker''} magnet which immediate consequently
removes the beam when the number of particles transcends 3000 per
spot. The spot area is about 6 $\mu$m$^2$ (beam
density of $\sim$$5\cdot 10^2$ nuclei/cm$^2$). The number of collisions
expected per spot is 17 \cite{EMU01-1}. The scanning efficiency
in the emulsion chamber is $0.75\pm 0.05$ \cite{EMU01-1}.

A detailed description of the configuration and the arrangement
of the emulsion chamber can be taken from \cite{TawDs}. Because
of the relatively high transverse momentum of produced particles
and fragmentations most of, if not all, these particles are
expected to be concentrated within narrow forward cone.
Therefore, their tracks can entirely be registered within the
forward emulsions coating the other seven plastic sheets. The
emulsion sensitivity for singly charged particles with minimum
ionization is found to be as good as 30 grains per 100 $\mu$m.
Depending on the incoming momentum, the polar angle can be
calculated. For the used data sample we get polar angle
$1.3$~mrad. Also, depending on the target foil and on the
topology of the microscope's field-of-view used for the image
acquisition \cite{ba242}, the measuring system can effectively
acquire secondary tracks with space angles $\theta$ up to
$30^{\circ}$  (pseudo-rapidity values $\eta =
-\ln\tan(\theta/2)> 1.32$). The singly charged particles are
basically expected to be mixed with small contamination of $e^-
e^+$ pairs coming from $\gamma$-conversions in the target foil and
from Dalitz production. As a reason of the reconstruction
algorithm applied in MIRACLE Lab \cite{ba242} the tracks of
these electrons cannot be grabbed. The possible {\it
overestimation} on the observed particles density is $\sim$2\%
\cite{TawDs}. In this value all the contributions of
$\gamma$-conversions in lead foil and in emulsion material have
been taken into account. The efficiency of MIRACLE Lab is up to
$96\%$. Large part of this 4\% discrepancy is obviously coming
from the tracks with relatively large space angle. In addition
the missing measurements, frequent scattering, unresolved close
pairs, and pair production of $e^- e^+$ represent an another
source of such discrepancy \cite{ba241}. Therefore, we could
consciously suggest to renounce the discussion of the effects of
$\gamma$-conversions on FM.

The investigation of FM is performed for Pb+Pb collisions with
particle multiplicity $> 1200$. From these central events only
the particles observed within predefined pseudo-rapidity
intervals are taken into account. These $\Delta\eta$ intervals
($3 < \eta < 4$, $3 < \eta < 5$, $2 < \eta < 5$, and $2 < \eta <
6$) are chosen at as well as around the mid-rapidity region of
interaction. The calculation of FM is performed by means of
Eq.\ (\ref{e:6}). As discussed before TPCF essentially get an
exponential increment while decreasing the relative momenta or
decreasing the pseudo-rapidity rather than a linear one. As
discussed in \cite{ba21,BecFM,TawPress2} TPCF are supposed to
contribute to the intermittent behavior of FM with considerable
values, especially for small relative momenta $Q$ or for small
phase-space partition of pseudo-rapidity $\delta\eta$. But as
contemplated in Section~\ref{bec_sec1} the identification of
secondary particles is highly required foremost to be able to
study TPCF in our experimental data. The identification of
produced particles by using the nuclear emulsion evidently is
not to put into practice \cite{ba241,ba242,TawDs}. Furthermore,
since there was no external magnetic field applied over the emulsion
chambers while the exposure process, measuring the particle momenta is
not to be implemented too. Although all these difficulties, I have
introduced in \cite{TawPress1} three different approximated
methods to adapt the calculations of TPCF to the data of nuclear
emulsion. For this destination the three algorithms are
utilized, in order to nearly identify the produced particles and
to approximately estimate their momenta.

\section{Definitions and Formalism} \label{bec_sec3}

\subsection{Two-particle correlation functions}
\label{bec_sec3.1}

By definition \cite{ba4,ba11,ba26}, the two-particle correlation
functions TPCF and their special case, the Bose--Einstein
correlations (BEC) are given as
\ba C_2(p_1,p_2) &=&
\frac{P_2(p_1,p_2)}{P_1(p_1)P_1(p_2)},\label{e:1}\\
C_2(Q^2)  &=& {\cal N} \left[1+\lambda
e^{-\frac{Q^2R^2}{2}}\right]. \label{e:2} \ea
$\lambda$ is the chaos parameter which essentially combines all
effects responsible for the chaotic phenomena taking place in
and affecting the entire dynamics of the emission source. For
example long resonance, coherent particle production, detector
resolution, contamination between non-pions and/or secondary
produced particles, etc. $P_1(p_i)$ is the probability to detect
the $i$-th single particle emitted with momentum $p_i$.
$P_2(p_i,p_j)$ is the probability to simultaneously detect the
$i$-th and $j$-th two particles emitted with momentum $p_i$ and
$p_j$, respectively. ${\cal N}$ is the normalization factor
depending only on the so-called {``event mixing''}
(Section~\ref{bec_sec3.3}). By means of the covariant one- and
two-particle distributions, the quantities $P_1(p_i)$ and
$P_2(p_i,p_j)$ can be expressed in dependence on the
multiplicity of one-particle $N_1(p_i)$ and two-particles
$N_2(p_i,p_j)$, respectively \cite{ba29}
\ba P_1(p_1) &=& E_1 \frac{d^3 n}{d^3 p_1}
                \equiv N_1(p_1), \label{e:3}\\
P_2(p_1,p_2) &=& E_1 E_2
                \frac{d^6 n}{d^3 p_1 d^3 p_2} \equiv
N_2(p_1,p_2).
\label{e:4} \ea
Then TPCF read the following expression
\be C_2(p_1,p_2)   = \frac{N_2(p_1,p_2)}{N_1(p_1)N_1(p_2)}.
\label{e:5}\ee
The experimental steps to estimate TPCF and afterwards to
calculate the parameters ${\cal N}$, $\lambda$, $R$ and/or
$\tau$ can briefly be summarized as follows:
\begin{enumerate}
\item Empirical estimation of $C_2$ by using the particle
  multiplicity. This estimation is simply the ratio of the correlated
  to the non-correlated particles as given in Eq.\ (\ref{e:5}).
\item Choosing an analytical expression for $C_2$, for instance
  Gaussian distribution which can be given in dependence on the
  dimension of emission source $R$ (Eq.\ (\ref{e:2})) or also on the
  life time $\tau$.  Depending on the modeling of the emission source
  density distribution we can extend these analytical expressions for
  different dimensions to retrieve estimation for other parameters.
  For the emulsion data I have driven a special expression given in
  Eq.\ (\ref{e:16}) \cite{TawPress1}.
\item Now the last step is more or less clear.  Fitting the
  experimental results of $C_2$ (step $\#$1) with the analytical ones
  (step $\#$2), to directly get estimation for ${\cal N}$, $R$,
  $\tau$, etc.
\end{enumerate}

\subsection{Factorial moments and intermittence exponents}
\label{bec_sec3.2}

If the phase-space for instance the pseudo-rapidity $\Delta\eta$
is split into $M$ bins of equal sizes $\delta\eta = \Delta
\eta/M$ and if $n_k$ being the multiplicity in $k$-th bin, then
the {``vertical''} FM are given as \cite{ba30}
\be <F_q(M)> = \frac{<n_k(n_k-1) \cdots  (n_k-q+1)>}{<n_k>^q},
\label{e:6}\ee
$<n_k>$ is the average multiplicity of the $k$-th bins of all
events. The self-similar density fluctuations for the successive
partitions leads to the following power-law scaling.
\be <F_q(M)> \propto M^{\phi_q}. \label{e:7}\ee
The parameters $\phi_q$ which represent
the slopes of the relations $\log F_q$ vs.\ $\log M$ are called
{``intermittence exponents''}. Their relations with the anomalous
dimensions \cite{ba31} are only depending on the orders $q$
\be d_q = \frac{\phi_q}{q-1}. \label{e:8}\ee
Therefore, for multi-fractal processes $d_q$ can simply be
given as functions of $q$ alone. $d_q=d_2$ characterizes the
mono-fractal processes, i.e.\ the particle production can be
characterized by one {\it fractal} dimension.

\subsection{Strip integral correlations}\label{bec_sec3.3}

The integrals of $q$-particle inclusive rapidity densities
$\rho_q(\eta_1,\eta_2,\cdots,\eta_q)$ are conditional to FM
given in Eq.\ (\ref{e:6}) \cite{ba32,ba33}
\be <F_q> =
\left<\frac{\int\limits^{\delta\eta}d\eta_1d\eta_2\cdots
d\eta_q~~\rho_q\left(\eta_1,\eta_2,\cdots,\eta_q\right)}{<n>^q}
\right>. \label{e:9} \ee
For the ideal case that there is no correlation between
produced particles $\rho_q(\eta_1,\eta_2,\cdots,\eta_q) =
\rho_1(\eta_1)\rho_1(\eta_2)\cdots\rho_1(\eta_q)$ and
consequently, $<F_q> \rightarrow 1$
\ba <F_q(\delta\eta)> &=&
\frac{\left<\int\limits_{\Omega_k}\prod\limits_{i=1}^q
      d\eta_i \;
\rho_q\left(\eta_1,\eta_2,\cdots,\eta_q\right)\right>}
{\hbox{{NORM}}}.
\label{e:10}\ea
$\hbox{NORM} = <\int_{\Omega_k}\prod_i^q d\eta_i \;
\rho_1(\eta_1)\rho_1(\eta_2)\cdots\rho_1(\eta_q)>$ is estimated
by the so-called {``event mixing''} \cite{ba32}. Generally, the
estimation of ``NORM'' plays a very important role with respect
to the final estimation of TPCF. In-offensively, one would
suggest to replace ``NORM'' in last equation by the distribution
of non-correlated particles. The difficulties to determine both
correlated and non-correlated particles in the emulsion data are
frequently discussed before. ``NORM'' is our data sample ---
non-correlated particle--pairs --- is estimated by using the three
methods given in \cite{TawPress1} (Section~\ref{bec_sec4.2.2}).
$<\int_{\Omega_k}\cdots>$ in Eq.\ (\ref{e:10}) represents the
correlated particle--pairs.

Using the strip integral correlations (Eq.\ (\ref{e:10})) instead of
the conventional sum over the bin multiplicity (Eq.\ (\ref{e:6}))
enables us to entirely avoid the large disadvantage known for
the second one, namely the artificial {\it fake} partitions
\cite{TawPress2}. In addition to these, the strip integral correlations
have other {\it important} advantages to decrease the statistical
errors and to increase the correlation size, etc. \cite{ba32}.
To be able to calculate the integral correlation functions in
the pseudo-rapidity phase-space $\Delta\eta$ obviously, we would
need to extend the strip integration to another domain,
individually depending on the pseudo-rapidity variable $\eta$
\be <C _q(\delta\eta)>=\frac{\left<\int\limits_{\Omega_{\eta}}
\prod\limits_{i=1}^q d\eta_i
     \rho_q(\eta_1,\eta_2,\cdots\eta_q)\right>}{\hbox{NORM}}.
\label{e:11}\ee
Besides the extension into $\eta$-domain performance the
counting of $q$-tuples\footnote[1]{The counting algorithm used in this article
has the following instructional iterations
\begin{itemize}
\item Determining the phase-space $\Delta\eta$ and the
      partition width $\delta\eta$. Within each phase-space
partition ($\delta\eta$),
      the differences between the pseudo-rapidity values
      $\eta_{ij}=\eta_i-\eta_j$; $i<j$ are calculated and then
      compared with the pre-specified width $\delta\eta$
according to the inner product if Eq.\ (\ref{e:13}).
\item Counting the $q$-tuple of $\{\eta_{ij_1}, \eta_{ij_2},\cdots,
      \eta_{ij_q}\}$ from each $i$-th event of the used data sample
      by using the {\it Heaviside} unit step function.
\item The results of last counting is then multiplied by $q!$
      to take into account the number of permutations within each $q$-tuple.
\item Repeating the last steps for all available \hbox{$(^n_q)-l$~tuples} and then
      calculating the average over the total events.
\end{itemize}
} in last equation represents the other
difficulty.  To avoid this, we use the
{Grassberger--Hentschel--Procaccia} integrals \cite{ba32,ba34}.
Therefore, the last strip integration for instance for the second
order ($q=2$) reads the following form
\be <C_2(\delta \eta)> = \frac{\left< 2! \sum\limits_{j_1<j_2}^n
\prod\limits_{k', k''}
    {\cal H} \left(\delta\eta - \left|\eta_{j_{k'}} -
    \eta_{j_{k''}}\right|\right)\right>}{\hbox{NORM}}.
\label{e:12}\ee
$n$ is the total particle multiplicity inside the whole considered
interval $\Delta\eta$ and $\cal H$ is the {``Heaviside''}
unit step function. Analogue to Eq.\ (\ref{e:12}) and by utilizing
suitable algorithm to count the strip integral correlations it
is possible to deduce a general expression for any order $q\ge2$.
This expression can also be given in dependence on the relative
pseudo-repidity $\delta\eta$ (Eq.\ (\ref{e:12})) or on the relative
momentum $Q$ (Eq.~(\ref{e:15})). For the direct relation between
both variables $Q$ and $\delta\eta$ we refer to Note~$\#$c
\be <C_q(Q^2)> = \frac{ \left< q! ~
     \sum\limits_{j_1<j_2<\cdots<j_q}^n ~ \prod\limits_{k',k''}^{q-1}
     {\cal H}\left(Q^2 - \delta Q^2_{j_{k'},j_{k''}}\right)\right>}{\hbox{NORM}},
\label{e:13}\ee
where $\delta Q^2_{j_{k'},j_{k''}}=\left|Q_{j_{k'}}-Q_{j_{k''}}\right|^2$.
The last relation is
perspicuously close coincidence with the expression of the
factorial moments given in Eq.\ (\ref{e:6}). By means of this
relation, we are apparently succeeded to derive an expression to
study the contributions of TPCF to the power-law scaling of the
conventional FM.

\subsection{TPCF as a possible source of intermittence}
\label{bec_sec3.4}

Once more the conditions of the validity of two-particle
interferometry are given in Section~\ref{bec_sec1}. Since the
breadth of the emission source is limited within the finite
source diameter $2R$, then TPCF (BEC) are effective, if
the relative momentum fulfill the following relation
\be \Delta p \le \frac{\pi \hbar}{R}\,. \label{e:14}\ee
In addition, from last expression the particle interferometry
(non-coherence phenomenon) strongly takes place, if the breadth
of the emission sources $R\rightarrow 0$. Of course we still
remember the properties of FM which are suggested as an
effective signature for the non-statistical fluctuations at the
final state of particle production \cite{ba30} and then as a
diagnostic tool to confirm the QGP formation. FM show power-like
behavior for the consecutive dividing of the phase-space
$\Delta\eta$ into $\delta\eta\rightarrow 0$ sub-intervals.
Equation (\ref{e:14}) describes almost the same behavior, namely
while the momentum difference is successively decreased $\Delta
p\rightarrow 0$, an exponential increasing of TPCF is strongly
expected. This behavior remains valid, even if the breadth of
emission source $R$ has a constant value (fixed source
diameter). In case that the dimension of emission source
sanctimoniously suffers a successive explosion or shrinking
$R\rightarrow 0$ it is obviously expected that TPCF become more
stronger. Practically, the {\it power-like} behavior of
two-particle correlation functions $C_2$ has been investigated
\cite{ba21} with the successive reduction of the momentum
difference $\Delta p\equiv Q\rightarrow 0$ as well as with the
successive reduction of the pseudo-rapidity interval
$\delta\eta\rightarrow 0$. This leads to the following power law
which apparently is to be compared with the scaling rule given in
Eq.\ (\ref{e:7}).
\be C_q\left(\Delta Q\right) \propto
\left(\Delta Q\right)^{-\phi_q} \label{e:15}\ee
TPCF are, as their name announces, the correlations between the
particle--pairs in a very restricted region ($\Delta r<6$~fm
and $\Delta p < 50$~MeV/c). Therefore, it is not
more difficult to explain why they are supposed to contribute to the
compulsory correlations between the produced particles?
Especially, if the investigation is being performed within the
region defined above, in which TPCF are effective. For all these
reasons it is now more or less clear, if we predict that TPCF
can be suggested as possible sources of the {``intermittent
behavior''} in heavy-ion collisions \cite{ba20,ba35}.

\section{Experimental Results} \label{bec_sec4}

\subsection{Results of factorial moments}\label{bec_sec4.1}

\begin{figure}[htb]
\vspace*{-5pt}
\begin{center}{\epsfxsize=8cm \epsfbox{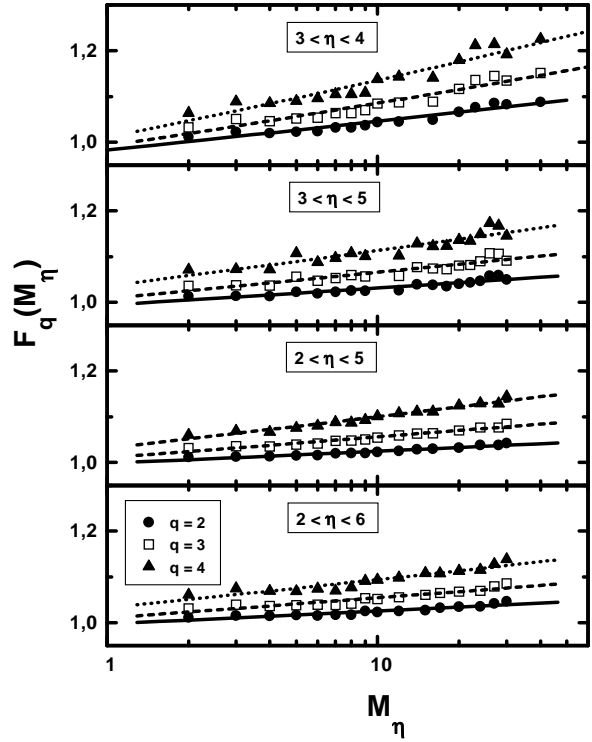}}
 \parbox{9cm}{\caption[]{\small\it The factorial moments $F_q$ for the orders $q=\{2,3,4\}$ are
  plotted as functions of {\it small} partition number $M_\eta$.  The
  data used here are Pb+Pb collisions at 158~AGeV with total
  multiplicity $>1200$. The results can good be fitted as straight
  lines (Eq.\ (\ref{e:7})). Their slopes represent the
  {``intermittence exponents''} $\phi_q$. Although the ordinates are
  also log scales, their labels are given.}\label{fig:1}}
\end{center}
\end{figure}

In all evens of the data sample the four pseudo-rapidity
intervals $\Delta\eta$ defined in Section~\ref{bec_sec2} are
successively split into $M\rightarrow \infty$ sub-intervals
(bins) with equal widths $\delta\eta=\Delta\eta/M\rightarrow
0$. In each such bins the particle multiplicity is counted and
then the $q$-order FM are calculated according to Eq.\ (\ref{e:6}).
Figure \ref{fig:1} shows $\log F_q$ as functions of $\log M_{\eta}$
for the orders $q=\{2,3,4\}$. The subscript in
$M_{\eta}$ means that the phase-space partition has been performed in
$\eta$-dimension only. Obviously, the experimental results can
good be fitted as straight lines according to the scaling rule
Eq.\ (\ref{e:7}). The slopes represent the so-called {\it
intermittence slopes} $\phi_q$. The four pictures show obvious
positive linear tendency. Also, it is to realize that the values
of $F_q(M_{\eta})$ and $\phi_q$ increase with decreasing the
$\Delta\eta$ intervals. The largest values are obtained within
the smallest interval $3<\eta<4$. There is another remarkable
finding that $\phi_q$ increase with increasing the orders $q$
({\it review} Eq.\ (\ref{e:6})).

\setlength{\textfloatsep}{10pt}

\setlength{\textfloatsep}{10pt}
\begin{figure}[htb]
\begin{center}{\epsfxsize=8cm \epsfbox{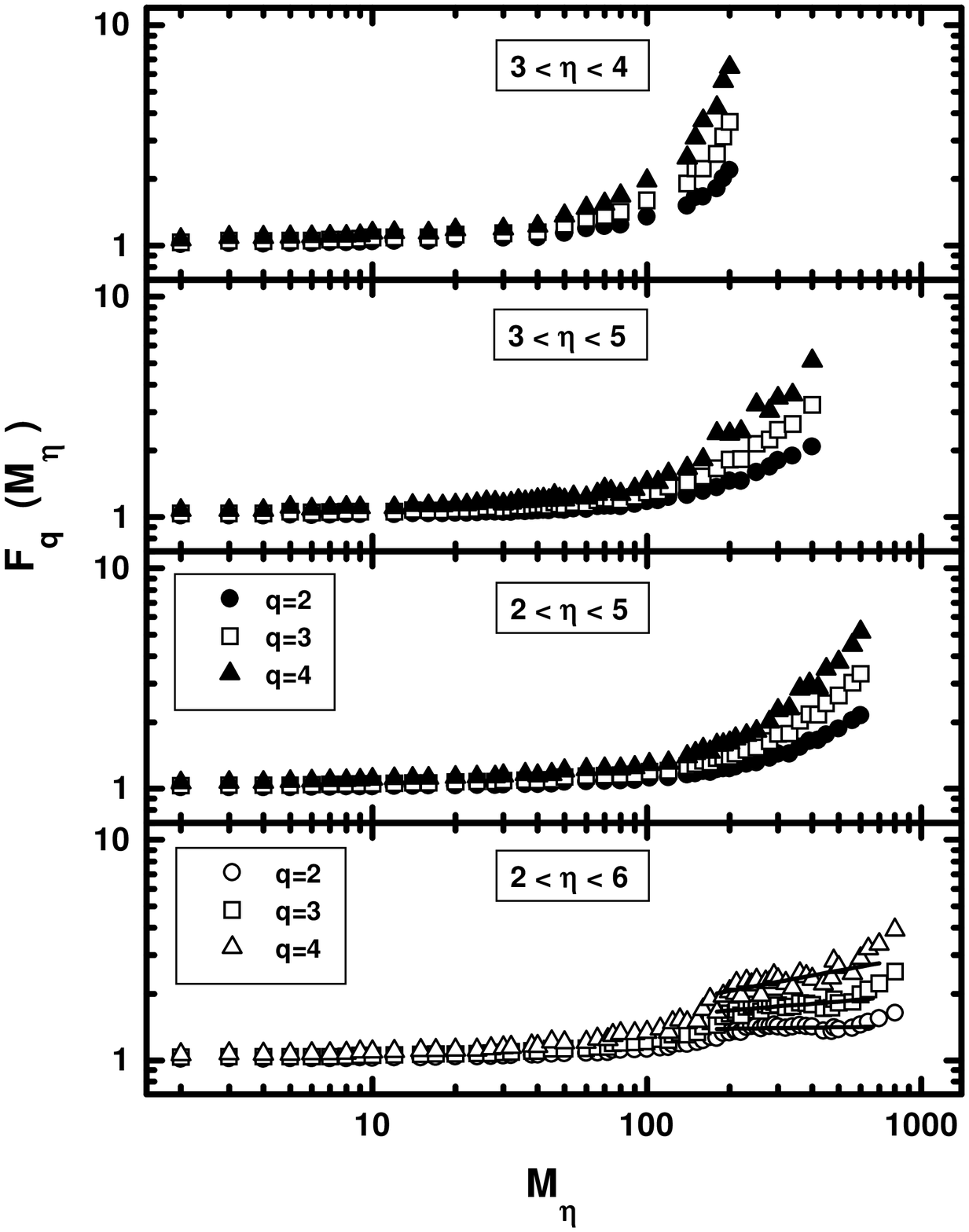}}
 \parbox{9cm}{\caption[]{\small\it The factorial moments $F_q$ for the orders $q=\{2,3,4\}$
  are depicted in log--log charts as functions of {\it large} partition
  number $M_{\eta}$. The partition process is repeated as long as
  FM values increase. An exponential growth of FM is noticed for
  relatively large $M_{\eta}$. Within the largest
  $\Delta\eta$ interval, we notice that FM are saturated. }
\label{fig:2}}
\end{center}
\end{figure}

By continuous increasing the partition number $M_{\eta}$, FM
exponentially grow (Fig.~\ref{fig:2}). The reasons for this fast
upwards increasing are discussed in details elsewhere
\cite{TawPress2}. Here the discussion is restricted only on the
phenomenological behavior of the increasing of $F_q(M_{\eta})$
with increasing the partition number $M_{\eta}$. Naturally,
increasing $\Delta\eta$ intervals enables us to divide the
one-dimensional phase-space ($\eta$) into large partition number
$M_{\eta}$. Once again we also notice in this figure that the
range of the linear dependence becomes larger with wider
$\Delta\eta$ windows (pictures from top to bottom). This linear
dependence are studied in the previous figure. In the top most
picture, the curvature upwards increasing has {\it sharp spikes}.
Plateau begins to appear with increasing the
$\Delta\eta$ intervals. In the largest interval (bottom most
picture) we have the largest plateau covering the distance from
$M_{\eta}\approx 200$ to $M_{\eta}\approx 600$. Beyond this
saturated regions $F_q(M_{\eta})$ continue their exponential
increasing. This time steeper than before. We continued the
partition process $\delta\eta \rightarrow 0$ as long as FM
increase.

\subsection{Results of TPCF} \label{bec_sec4.2}

\subsubsection{Adapting TPCF in emulsion data}
\label{bec_sec4.2.1}

As shortly introduced before it is expected that TPCF contribute
to FM with considerable values, especially, if FM are studied
within small phase-space intervals where TPCF are expected to be
strongly enhanced (Section~\ref{bec_sec1}). Therefore, before the
announce whether non-statistical fluctuations (e.g.\ dynamical)
are observed as the case in Fig.~\ref{fig:1} and \ref{fig:2},
one has to study all possible sources which may be responsible
for such {\it abnormal} scaling behavior. After that if the
further scaling of FM still establish linear positive tendencies
with the partition number, this might be then referred to
self-similar branching processes and/or to critical second-order
phase transition to QGP \cite{TawPress2,TawNew}. Regard being had
to the nature of TPCF alone, one can expect an exponential
dependence with the decreasing of the considered phase-space
rather than a linear one ({\it review} Eq.\ (\ref{e:2})). As
mentioned in Section~\ref{bec_sec1} the identification of
produced particles and the estimation of their momenta are
highly necessary to be able to study the two-particle
interferometry. The two-particle interferometry known as BEC
(Section~\ref{bec_sec1}) represents the simplest case of TPCF.
The identification of produced particle by using nuclear emulsion
obviously is not to be put into practice. In \cite{TawPress1}
three different methods have been introduced and examined to
make it possible to estimate TPCF for the emulsion data sample.
For these algorithms the expression given in Eqs (\ref{e:2}) and
(\ref{e:5}) must be modified as follows:
\be C_2^{Em}(Q) = \frac{1}{2} \left[1 + {\cal N}\left(1+ \lambda
e^{-\frac{Q^2R^2}{2}}\right)\right]. \label{e:16}\ee
The prediction of this approximation are compared with the ratios
between correlated and non-correlated multiplicity of particle--pairs
(Eq.\ (\ref{e:5})). The superscript in Eq.\ (\ref{e:16}) says that the
investigation and consequently, this analytical form are applied for
emulsion data sample. The subscript refers to the two-particle
correlations BEC. This will be replaced by $q$, when we will modify
Eq.\ (\ref{e:16}) to be compared with FM for the different orders
$q$. The results of Eq.\ (\ref{e:16}) will be depicted in
Fig.\ \ref{fig:3}.  The fitting parameters ${\cal N}$, $R$, and
$\lambda$ are compared with the parameters of data samples acquired
by other detectors which are in the position of being able to identify
the produced particles and to measure their impulses \cite{opal1}.

\subsubsection{TPCF as function of $Q$}\label{bec_sec4.2.2}

\begin{figure}[htb]
\begin{center}{\epsfxsize=8cm \epsfbox{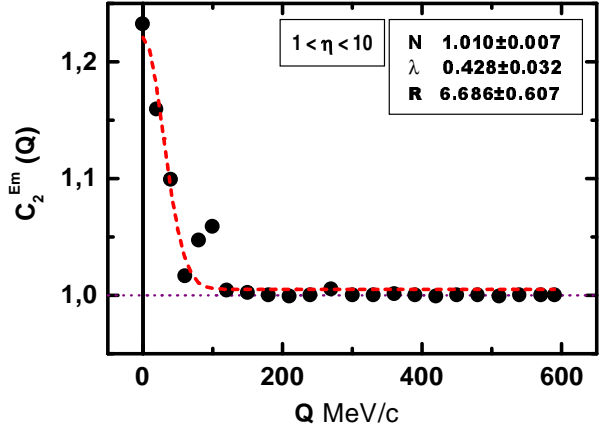}}
 \parbox{9cm}{\caption[]{\small\it $C^{Em}_2(Q)$ are depicted as functions of the
invariant relative momentum $Q$ for the interval $1<\eta<10$.
The fitting parameters obtained by using the analytical relation
Eq.~(\ref{e:16}) are given at the upper right corner.}
\label{fig:3}}
\end{center}
\end{figure}

Figure \ref{fig:3} shows the dependence of TPCF on the invariant
relative momentum $Q$ of all possible particle--pairs of our
Pb+Pb data sample observed within the pseudo-rapidity interval
$1<\eta<10$. For this data set we counted about
$\sim$$3\cdot10^{6}$ particle--pairs. The averaged {\it transverse}
momentum $<p_t>$ is assumed to be constant $\sim$350~MeV/c. By
using this constant value and the related equations given in
\cite{TawPress1} it is possible to assign to each particle a
stochastically calculated momentum. Then it is possible to
calculate the relative momentum $Q$ for each particle pair. The
algorithms discussed in \cite{TawPress1} are used to count the
non-correlated particle--pairs and then to classify their
multiplicities according to $Q$. Counting the correlated particle
pairs is now more or less unsophisticated process. In fact the
algorithm which randomizes the angular characters of the original
events, is preferentially used. Obviously, the randomized
particles in this regard represent the background distribution
``NORM''. In such a way the normalization factor ${\cal N}$
gets an average value fixed around the unity. It is necessary to
point out here that no correction for the Coulomb repulsion has
been considered, since the amount of Gamov correction factor
\cite{ba52} is $\sim$10\% for $Q\approx 1$~MeV/c. With increasing
the values of $Q$ Gamov correction factor exponentially decays.
As discussed in Section~\ref{bec_sec1} the two-particle
interferometry must be taken into account only for
$Q\le50$~MeV/c. Therefore, the ratios of correlated particle
pairs (multiplicity of original events) to the non-correlated
ones (multiplicity of background distributions) for $Q
>70$~MeV/c can be fixed to the unity, i.e.\ for $Q
>70$ MeV/c, we can assume that there is no difference between
the multiplicity of correlated and non-correlated particle--pairs.

\setlength{\textfloatsep}{10pt}

The experimental results shown in Fig.\ \ref{fig:3} are fitted by
using Eq.\ (\ref{e:16}). The obtained fitting parameters are
$R=6.686\pm0.607$~fm, $\lambda=0.428\pm0.032$, and ${\cal
N}=1.010\pm0.007$. All these values are comparable with the
results obtained from data sample of detectors able to identify
the produced particles and to estimate their charges and momenta
\cite{opal1}. For our one-dimensional analysis the resulting
parameters are assumed to combine all their corresponding
components. Using one-dimensional analytical expression for the
correlation functions $C$ as in Eq.~(\ref{e:16}) which depends
only on the invariant relative momentum $Q$ obviously does not
enable us to directly investigate the behavior of the other
components of $R$ and $\lambda$ with $Q$. The breadth of
emission source $R=6.686\pm0.607$~fm is on the one hand
evidently comparable with the r.m.s.\ nuclear radius of $^{82}$Pb,
$7.11$~fm on the other hand it shows that the method used here to
calculate BEC in emulsion data sample has a very good physical
acceptance. The value of the chaos parameter $\lambda<1$ is an
evidence for the partial non-coherent emission source and for
the simultaneously emitted particles. As given in
Section~\ref{bec_sec3.1} this parameter includes all possible
chaotic effects taken place in the emission source. More details
about the results of BEC in the emulsion data sample can be
obtained from \cite{TawPress1}.

\subsubsection{TPCF as function of $\eta$}\label{bec_sec4.2.3}

The conventional TPCF dependence on the invariant relative
momentum $Q$ does not enable us to directly\footnote[2]{
Since the invariant relative momentum $Q$ on the one
hand depends on the squared invariant mass
$Q^2=M_{inv}^2-(im)^2$, it can be used to study the
fluctuations due to cluster resonance decay. On the other hand this
relation reads
\ba Q^2&=&[p_{1t}^2 + p_{2t}^2 +2p_{1t}^2 p_{2t}^2
\cos(\delta\phi)] \nonumber \\
&-&[M_{1t}^2 + M_{2t}^2 +2M_{1t}^2 M_{2t}^2
\cos(\delta\eta)]\,.\nonumber \ea
This relation implies that for given
$\delta\phi=|\phi_1-\phi_2|$, $Q=|p_1-p_2|$ can be given in
dependence on $\delta\eta=|\eta_1-\eta_2|$. Such a way TPCF can
also be studied in dependence on the phase-space partition of
pseudo-rapidity $\delta\eta$.
} study their
contributions to the intermittent behavior observed in the
experimental data (Fig.\ \ref{fig:1} and \ref{fig:2}). For this
purpose, it is needed to study as evident as possible the
direct TPCF dependence on the pseudo-rapidity $\eta$ or
alternatively to investigate their behavior within the region,
$Q<50$ MeV/c. Such a way it will be possible to check whether
their behavior can be given by a power law similar to that of FM,
i.e.\ whether TPCF are also intermittent.
Section~\ref{bec_sec4.2.4} is devoted to the second suggestion.
The behavior for large interval $0.5<Q<100$~MeV/c will be
studied and compared with the results given in
Section~\ref{bec_sec4.1}. In this section we study the direct
dependence of $C^{Em}_2$ on the pseudo-rapidity phase-space
$\delta\eta$.

\setlength{\textfloatsep}{10pt}
\begin{figure}[htb]
\begin{center}{\epsfxsize=8cm \epsfbox{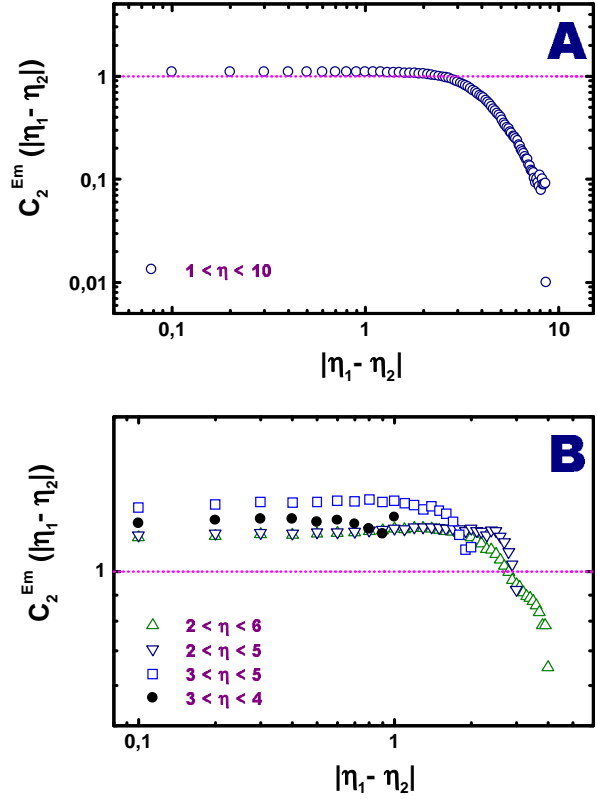}}
 \parbox{9cm}{\caption[]{\small\it The dependencies of 
$C^{Em}_2(|\eta_1-\eta_2|)$ on the $\eta$ differences
$|\eta_1-\eta_2|$ are drawn in {A}. $C^{Em}_2$ have values
greater than $1$ for the $\eta$ differences $<2.0$. After that
their values decrease exponentially. The same relations are drawn
in {B} for the other $\Delta\eta$ intervals. The linear
dependencies are also still to be noticed here. Also the
exponential decrease for $\delta\eta>2.0$.} \label{fig:4}}
\end{center}
\end{figure}

Figure \ref{fig:4}A shows the dependence of $C^{Em}_2$ on
$\delta\eta$ for the pseudo-rapidity interval $1<\eta<10$. The
intervals between the successive iterations of $\delta\eta$'s are
constant ($0.1$). Taking the different statistical acceptances
into account we notice that the step-wise successive increasing
of $\delta\eta=|\eta_1-\eta_2|$ leads to slow {\it linear}
decreasing of $C^{Em}_2$ until the value $\delta\eta\approx
2.0$. This value is corresponding to space angles $\theta\approx
15^{\circ}$, i.e.\ TPCF of particle--pairs distributed up to
$\sim$$15^{\circ}$ are able to show a power-like scaling. Almost
the same behavior we observed in Fig.\ \ref{fig:1}. It is
clear that $\delta\eta=|\eta_1-\eta_2|$ used here does not
directly coincident with the bin size $\delta\eta=\Delta\eta/M$
used in Fig.~\ref{fig:1} and \ref{fig:2}. This linear decreasing
with increasing $\delta\eta$ obviously characterizes the expected
scaling rule. Therefore, it can be compared with that of the FM
(Eq.\ (\ref{e:7})). As given before the intermittence slopes
$\phi_q$ can be directly retrieved from such log--log scale. If
we continue to increase $\delta\eta$, the decrease of $C^{Em}_2$
becomes exponential. And TPCF become smaller than 
unity.\footnote[3]{
The ratios between the correlated and the non-correlated
         particle--pairs define the two-particle correlation functions
         $C_2$. For the value $C_2<1$ we can suppose that the ratios and
         the
         corresponding functions are not valid any more.
} For
very large $\delta\eta$ some kind of plateau (saturated regions)
are formed. In Fig.~\ref{fig:3} we showed that TPCF or BEC are
applicable only for $Q$ up to $\sim$50~MeV/c. Figure~\ref{fig:4}A
can be used to visualize the conclusion that the functions
$C^{Em}_2(\delta\eta)$ can be compared with $C^{Em}_2(\Delta
Q)$. We can therefore conclude that the functions
$C(\delta\eta)$ are to be used for $\delta\eta<2.0$ and
eventually, for $\Delta Q<50$~MeV/c. That the functions
$C_2(\delta\eta)$ have linear dependence on the $\delta\eta$
differences until $\delta\eta\approx 2.0$ leads to the
assumption that until this value we can describe their
dependencies by using a power-law scaling. Then the two values
$\delta\eta\approx 2.0$ and $\Delta Q\approx 50$~MeV/c are
coincident with each other. They apparently represent
the upper limit to consider the TPCF effects. Beyond these
values TPCF do not affect the observed FM.

The bottom picture (Fig.~\ref{fig:3}B) depicts the same relations with the same
step-width $0.1$ but for the other $\Delta\eta$ intervals. The
linear dependencies depending on the $\Delta\eta$ intervals\footnote[4]{
This generally valid for all $\Delta\eta$ intervals.
For the first two large intervals as well as for the interval $1<\eta<10$,
evidently this value is included. For the other two intervals it
is excluded, since the differences $|\eta_1-\eta_2|$ are not
enough to reach this value.
} are also still to be noticed here. They expand also until the
value $\delta\eta\approx 2.0$. Although, $C^{Em}_2(\delta\eta)$
for the small $\Delta\eta$ intervals leave the linear dependence
earlier than for the large ones evidently, almost the same
behavior is also noticed here. There is an interesting finding
that the increasing of $\delta\eta$ differences results a giant
decreasing of the correlation function $C^{Em}_2(\delta\eta)$.
In this regard we can conclude that the effects of TPCF
exponentially weakened with increasing the differences
$\delta\eta$. Especially, within the two large intervals
$2<\eta<5$ and $2<\eta<6$ we have similar saturated regions as
in the top picture. So far we can summarize that the comparison
between this figure and Fig.\ \ref{fig:1} and \ref{fig:2} shows
that our investigation of TPCF in dependence on $\delta\eta$
differences results essential information about their
contributions to the intermittent behavior.

\subsubsection{Strip integral correlations} \label{bec_sec4.2.4}

\begin{figure}[htb]
\begin{center}{\epsfxsize=8cm \epsfbox{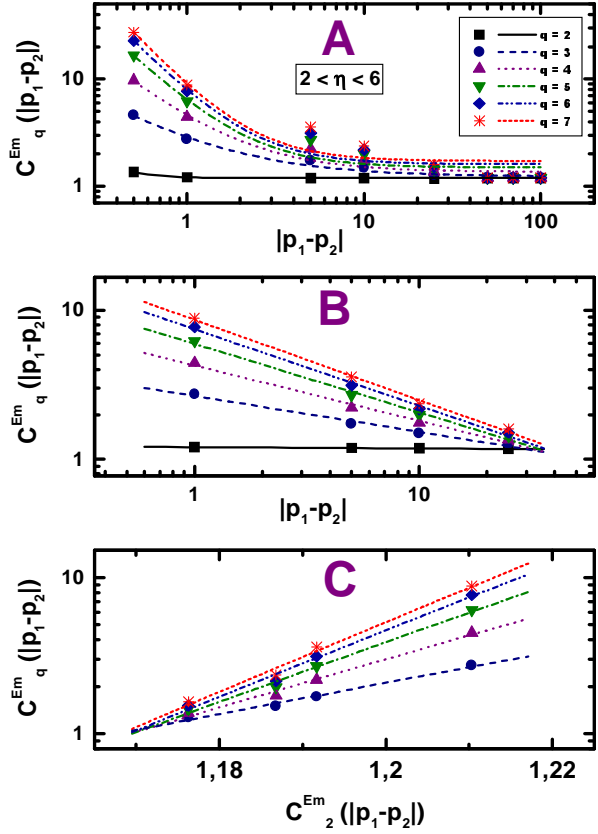}}
 \parbox{9cm}{\caption[]{\small\it For the orders $q=\{2,3,4,5,6,7\}$ the
  strip density integrals $C^{Em}_2$ are drawn in {A} as
  functions of the distances $|p_1-p_2|$. The experimental results
  are fitted by using scaling rule given by Eq.\ (\ref{e:17}). The fitting
  parameters are given in Table~\ref{tab:1}. The same relations are
  depicted in {B} for the region $1\le |p_1-p_2| < 50$~MeV/c.
  These can good be fitted as straight lines (Eq.\ (\ref{e:15})).
  Their fitting parameters are listed in Table~\ref{tab:2}. The
  relations between the $q$-order $C^{Em}_q$ and the second-order
  ones are illustrated in {C}.  The results are fitted according
  to Eq.\ (\ref{e:18}). The fitting parameters are listed in
  Table~\ref{tab:3}. Although we also use here a log--log scale, the
  labels of the abscissa are given.}}
\label{fig:5}
\end{center}
\end{figure}

Using the strip integral correlations given in
Section \ref{bec_sec3} we can calculate FM in dependence on the
invariant relative momentum $Q$. This represents an extension of
the second alternative method suggested in
Section~\ref{bec_sec4.2.3} to check whether the behavior of TPCF
for the orders $q=\{2,3,4,5,6,7\}$ can be given by scaling
rules similar to that of FM. The results of strip integral
correlations are given in Fig.~\ref{fig:5}A for the interval
$2<\eta<6$. Starting with a certain momentum difference we
compare it with the calculated relative momentum $|p_1-p_2|$ by
means of the {\it Heaviside} unit step function ${\cal H}$ as
given in Eq.\ (\ref{e:13}). We can clearly realize that the
increasing of the momentum differences, i.e.\ increasing the
distances between the particles, within which we are searching
for the possible particle--pairs which fulfill the condition
$Q^2-(|p_1-p_2|^2)>0$ results an additional decreasing of FM.
Taking into account the effects of the inner product in
Eq.\ (\ref{e:13}) the proportional increasing of FM according to the
orders $q$ can be understood.

\setlength{\textfloatsep}{10pt}

In light of the scaling Eq.\ (\ref{e:15}) we study the dependence
of $C^{Em}_q(\Delta Q)$ on $\Delta Q$ for relative momenta
ranging from $0.5$ until $100$~MeV/c. Obviously, we notice that the
ability of Eq.\ (\ref{e:15}) to describe these results is limited.
Therefore, it should be modified to simulate these {\it
non-linear} relations. The validity of Eq.\ (\ref{e:15}) will be
discussed later. We would therefore suggest the following
empirical power law
\be C_q^{Em}(\Delta Q) = a + b \cdot (\Delta Q)^{-c}\,.
\label{e:17}\ee
Using this scaling rule the experimental results in
Fig.\ \ref{fig:5}A are fitted and the resulted parameters given in
Table~\ref{tab:1}. We generally notice that for all relations
the fitting parameters increase with increasing the orders $q$.
There is only one exception for $q=2$, $c$ is too large
meanwhile $b$ is too small. By means of this scaling rule it is
not possible to retrieve any information about the contribution
of TPCF to the observed FM, since the intermittence slopes are
entirely hidden. For this destination we should first divide the
region $0.5<\Delta Q<100$~MeV/c into sub-intervals possibly
with common properties. As discussed before TPCF have a
restricted region outside of it they are not applicable. Last
figure serves to view these properties. Visually, one can notice
for instance for $\Delta Q>50$~MeV/c the values of $\log
C_q^{Em}$ are constant for all $q$.

Although we could not retrieve more experimental points for the
region $\Delta Q<1$~MeV/c, it is to realize that within this
region $C^{Em}_q(\Delta Q)$ exponentially decrease with
increasing the $\Delta Q$ differences. Although this decrease
continues even until \hbox{$\Delta Q \approx 50$~MeV/c}, we can
notice that the decrease within the region $1\le\Delta
Q<50$~MeV/c is not exponential. We still remember the properties
of the functions $C^{Em}_q$ within this restricted region. They
are comparable with the factorial moments given in
Section~\ref{bec_sec3.2}. As we noticed they showed an obvious
intermittent behavior. Therefore, the subscript $q$ gives the
order of FM. For \hbox{$\Delta Q>50$~MeV/c} we can assure that
$C^{Em}_q(\Delta Q)\approx 1.0$ for all $\Delta Q$, i.e.\ TPCF
have no more effects on the observed particle correlations, if the
two-particle relative momenta are greater than $50$~MeV/c. These
facts are visualized in this figure. We obviously notice that all
experimental point are saturated.

To study the behavior within the region $\Delta Q<1$ the
experimental data should be first percolated from the Coulomb
repulsion. The Gamov correction factor which is usually used to
measure the possible interactions between non-coherently emitted
particles, i.e.\ particles simultaneously emitted with almost the
same momenta, is about $10\%$ for $\Delta Q\approx 1$~MeV/c. It
decreases exponentially with increasing the relative momenta
$\Delta Q$. It has a negligible value for few MeV/c. The
experimental results used here are not liberated from these
interactions. Therefore, we can exclude this small region while
studying the dependence of $C^{Em}_q$ on $Q$.

\begin{table}[htb]
{\baselineskip=11pt
\caption{\small\it The relations between $\log C(\Delta Q$) and $\log\Delta Q$
drawn in Fig.\ \ref{fig:5}A for $0.5\le\Delta Q<100$~MeV/c are
fitted according to Eq.\ (\ref{e:17}). The fitting parameters
$a$, $b$, and $c$ are listed here. We notice that all values increase
with increasing the orders $q$. \baselineskip=10pt}\label{tab:1}}
\begin{center}
\begin{tabular}{ c c c c } \hline\\[-10pt]
$q$ &      $a$        &       $b$       &      $c$ \\[2pt] \hline\\[-10pt]
2   & $1.188\pm0.002$ & $0.022\pm0.006$ & $2.882\pm0.984$ \\
3   & $1.225\pm0.066$ & $1.654\pm0.145$ & $1.004\pm0.116$ \\
4   & $1.365\pm0.136$ & $3.263\pm0.339$ & $1.335\pm0.144$ \\
5   & $1.496\pm0.211$ & $4.938\pm0.546$ & $1.607\pm0.156$ \\
6   & $1.606\pm0.277$ & $6.405\pm0.723$ & $1.721\pm0.161$ \\
7   & $1.721\pm0.349$ & $7.412\pm0.913$ & $1.792\pm0.175$ \\
\hline
\end{tabular}
\end{center}
\end{table}

\begin{table}[htb]
{\baselineskip=11pt \caption{\small\it Within the region
$1\le\Delta Q<50$ the relations $\log C(\Delta Q$) vs.\
$\log\Delta Q$ depicted in Fig.\ \ref{fig:5}B can good be fitted by
Eq.\ (\ref{e:15}). We notice that the intermittence slopes $\phi_q$
increase with the increasing orders $q$. The constants $k$
represent the intersects. \baselineskip=10pt}\label{tab:2} }
\begin{center}
\begin{tabular}{c c c } \hline\\[-10pt]
$q$ & k & $\phi_q$ \\[2pt] \hline\\[-10pt]
2 & $0,083\pm0,001$      & $0,009\pm0,001$ \\
3 & $0,426\pm0,018$      & $0,241\pm0,020$ \\
4 & $0,631\pm0,024$      & $0,372\pm0,026$ \\
5 & $0,776\pm0,026$      & $0,461\pm0,026$ \\
6 & $0,874\pm0,026$      & $0,514\pm0,028$ \\
7 & $0,937\pm0,020$      & $0,539\pm0,022$ \\
\hline
\end{tabular}
\end{center}
\end{table}

Thence, we are left with the region $1\le\Delta Q<50$~MeV/c,
only. For this region the dependence of $C^{Em}_q(|p_1-p_2|)$
on  the relative momenta $\Delta Q=|p_1-p_2|$ are drawn in
Fig.\ \ref{fig:5}B. We notice that the relations in this log--log
scale can good be fitted as straight lines (Eq.\ (\ref{e:15})). This
behavior is entirely comparable with all pictures in
Fig.\ \ref{fig:1} and in the linear regions in Fig.\ \ref{fig:4}. As
given before the tendencies represent the {\it intermittence
slopes} $\phi_q$ which are listed in Table~\ref{tab:2}. The
increasing the orders $q$ the increasing the slopes $\phi_q$.
This behavior characterizes the intermittent interacting system.
Within this region we can then quantitatively estimate the
contributions of TPCF (BEC) to the corresponding FM. To do this
we should first study the dependence of the ratios
$\phi_q/\phi_2$ on the orders $q$. Motivated by the Ochs--Wosiek
log--log scaling \cite{ba44} by the adoption of $C^{Em}_q$ to the
conventional FM given in Section~\ref{bec_sec3.2} and by the
limited validity of Eq.\ (\ref{e:15}) we would therefore recommend
the following power-law scaling to be applied within the region
$1\le\Delta Q<50$~MeV/c
\be \log C^{Em}_q(\Delta Q)= \frac{\phi_q}{\phi_2}\cdot\log
C^{Em}_2(\Delta Q)+k\,.
\label{e:18}
\ee

\begin{table}[htb]
{\baselineskip=11pt \caption{The dependencies of
$q$-order $C^{Em}_q(\Delta Q)$ on the second-order ones
depicted in Fig.\ \ref{fig:5}C for the restricted region
$1\le\Delta Q<50$~MeV/c are fitted according
to Eq.\ (\ref{e:15}). We clearly notice that the ratios
$r_q\equiv\phi_q/\phi_2$ increase with increasing the orders
$q$. Defining the variables $m_q=r_q-(^q_2)$ and
$n_q=r_q-(q-1)$ we compared them with $r_q$. Also
the differences between the successive orders of $m$ and $n$
are given. \baselineskip=10pt}\label{tab:3}} 
\begin{center}
\begin{tabular}{ c c c c  c c c c} \hline\\[-10pt]
$q$ & $r_q$ & $-k$ &  & $m_q$ & $m_q-m_{q-1}$ & $n_q$ & $n_q-n_{q-
1}$\\[2pt] \hline\\[-10pt]
3   & $27,52\pm2,18$ & $1,85\pm0,17$ & & $24$ & $23$ & $25$ & $24$ \\
4   & $42,37\pm2,99$ & $2,88\pm0,22$ & & $36$ & $12$ & $39$ & $14$ \\
5   & $52,43\pm3,63$ & $3,56\pm0,27$ & & $42$ & $6$  & $48$ & $9$  \\
6   & $58,43\pm4,03$ & $3,96\pm0,30$ & & $43$ & $1$  & $53$ & $5$  \\
7   & $61,17\pm4,33$ & $4,13\pm0,33$ & & $40$ &        & $55$ & $2$  \\
\hline
\end{tabular}
\end{center}
\end{table}
\vfill\pagebreak

The relations between the $q$-order $C^{Em}_q(\Delta Q)$ and the
second-order ones are given in Fig.\ \ref{fig:5}C. As expected all
these log--log relations have positive linear dependencies where
the ratios of intermittence slopes $\phi_q/\phi_2$ can directly
be retrieved from these relations. $\phi_q/\phi_2$ provide
important information about the reaction dynamics and also about
the properties of intermittent behavior
\cite{TawDs,TawPress2,TawNew}. The fitting parameters according
to the scaling rule Eq.\ (\ref{e:18}) are listed in
Table~\ref{tab:3}. We notice that $\phi_q/\phi_2$ apparently
increase with increasing the orders $q$. These ratios are too
large in comparison with $(^q_2)$ which are related to the {\it
self-similar process}. They are also larger than the values
resulting from the scaling rule of critical phenomenon
characterized by $(q-1)$. For the same data sample we could
proof on the one hand that the intermittence slopes
$\phi_q/\phi_2$ cannot be simulated as {\it self-similar
process} \cite{TawPress2,TawNew}. Their values are smaller than
$(^q_2)$. On the other hand we could partially fit the
experimental results according to the critical phenomenon
$(q-1)$. In light of these results we try in next section to
give an answer to the question {``How much can TPCF
contribute to the observed FM?''}. We will measure their
contributions to the most important parameter of the intermittent
behavior, namely, the ratios $\phi_q/\phi_2$.

\subsubsection{TPCF contributions to FM} \label{bec_sec:4.2.5}

We define $r_q\equiv\phi_q/\phi_2$, $m_q=r_q-(^q_2)$, and
$n_q=r_q-(q-1)$. The differences between the successive orders of
$m$ and $n$ are also calculated and listed in
Table~\ref{tab:3}. Taking into account the assumption that the
contributions of TPCF are linearly added to the observed FM.
Furthermore, we know that the experimental values of
$\phi_q/\phi_2|_{\rm exp}$ obtained from FM are calculated over wide
ranges of pseudo-rapidity phase-space $\Delta\eta$
(Section~\ref{bec_sec4.1}). Then we can suggest to extrapolate them
into the relatively small restricted region of the valid TPCF
($\delta\eta<2.0$ or $\Delta Q< 50$~MeV/c). Therefore, we  can
suppose that the values of $\phi_q/\phi_2|_{\rm exp}$ are still valid
within the region $1\le\Delta Q<50$~MeV/c. In the language of
pseudo-rapidity we suppose the validity of
$\phi_q/\phi_2|_{\rm exp}$ for $\delta\eta<2.0$. As given above both
$(^q_2)$, and $(q-1)$ could be used on the one hand to
simulated the experimental results of FM
($\phi_q/\phi_2|_{\rm exp}$). On the other hand both of them are
unable to fit the results of $r_q$. We noticed from
Table~\ref{tab:3} that $r_q$ are larger than both $(^q_2)$ and
$(q-1)$. According to these assumptions we can suppose to
subtract $\phi_q/\phi_2|_{\rm exp}$ from $r_q$ observed within the
region of effective TPCF. Such a way we get a primitive
estimation of TPCF contributions to FM. The increasing the
orders $q$ the increasing both $m_q$ and $n_q$. For $(^q_2)$
and $(q-1)$ we can realize that the differences between the
successive orders increase with increasing the orders $q$. In
contrast we notice for $r_q$ and correspondingly for both $m_q$
and $n_q$ that the differences of their successive orders
decrease with increasing $q$. For large orders we expect that
$m_q=m_{q-1}$ and $n_q=n_{q-1}$. This can be referred to the
limited rapidity values of the used data sample and to the
restricted available dimension of the emission source.

\section{Summary and Final Conclusions} \label{bec_sec5}

In this article we have studied the non-statistical fluctuations
in Pb+Pb collisions at $158$~AGeV. First, we would like to sum up
the results obtained so far. The analysis of FM in one-dimension
shows that the interacting system is obviously intermittent. The
phase-space partitions are performed as long as FM increase. For
large partition number the increase of FM is exponentially
binding upwards. Part of this fast increasing can be understood
as TPCF effects. For large $\Delta\eta$ intervals (wide data
sample), saturated regions are formed. By continuous increasing
the partition number FM continue to rise fast. The difficulties
to apply BEC in the emulsion data are shortly discussed.
Nevertheless we applied one of the methods suggested in
\cite{TawPress1}. Using a special analytical expression for TPCF
we could retrieve estimations for the dimension of emission
source $R$ and for the chaos parameter $\lambda$. As expected
the randomization parameter $\lambda\approx 1$. This values is
an evidence for partially non-coherent emission source and for
the simultaneously emitted particles. For the dimension $R$ the
value we get is comparable with the r.m.s.\ $^{82}$Pb radius. These
results further the use of our method to estimate BEC in the
emulsion data sample. This enabled us to directly pass to the
main destination of this article. The study of the TPCF
contribution to the observed FM. First we studied the dependence
of TPCF on the phase-space of pseudo-rapidity (on the
differences, $\delta\eta$). For small $\delta\eta$ differences
we get linear dependencies for all $\Delta\eta$ intervals until
$\delta\eta\approx 2.0$. After that the dependence becomes
exponential. This limitation is coincident with the results shown
in Fig.\ \ref{fig:3}. An important finding found here is that the
boundary between the regions of effective and non-effective TPCF
is located around $\delta\eta\approx 2.0$. For relative momenta
the boundary is located around $\Delta Q\approx 50$~MeV/c.
Defining the region of effective TPCF in $\delta\eta$ enabled us
to directly study their intermittent behavior. Using the strip
integral correlations we have driven a relation for the TPCF
dependence on the phase-space pseudo-rapidity $\delta\eta$ in
regard to the intermittent behavior. The dependence of $\log
C^{Em}_q(\Delta Q)$ on $\log\Delta Q$ for the region $1\le\Delta
Q<50$~MeV/c supports our conclusion that TPCF show an
intermittent behavior and probably add considerable values to
the conventional FM. The dependence of the intermittence slopes
$\phi_q$ on the orders $q$ can confirm the conclusion
concerning the intermittent behavior. To quantitatively estimate
the contributions of TPCF we first studied the dependence of
$q$-order $C^{Em}_q$ on the second-order one. Here we noticed
that the characterized behavior of intermittency, namely the
ratios of intermittence slopes $\phi_q/\phi_2$ increase with
increasing the orders $q$. These ratios are compared with
results given in \cite{TawPress2,TawNew}. It is found that on
the one hand the values supposed to be added by TPCF increase
with increasing the orders. On the other hand the differences
between their successive values decrease with increasing orders.

\section*{Acknowledgements}

I am very grateful to E.~Stenlund and all colleagues of EMU01
Collaboration for the helpful discussions and the kind
assistance. Especially, that they allowed the use of part of our
collaborative experimental data for this work. I would like to
thank E.~Ganssauge for the continuous support, the interesting
discussions, and the helpful comments. It is my pleasure to thank
to \hbox{C.~Radke} for the careful reading of the manuscript.

\end{document}